\documentclass[10pt,twocolumn]{article}

\usepackage[utf8]{inputenc}
\usepackage{amsmath}
\usepackage{amssymb}
\usepackage{amsthm}
\usepackage{titlesec}
\usepackage{cite}
\usepackage{color}
\usepackage{booktabs}
\usepackage{graphicx}
\usepackage{nicefrac}
\usepackage[colorlinks=false]{hyperref}
\usepackage[format=plain,labelfont=it]{caption}
\usepackage[left=1.5cm,right=1.5cm,top=2cm,bottom=2cm]{geometry}
\usepackage{algpseudocode}
\usepackage[font=footnotesize]{caption}

\renewcommand{\thesubsubsection}{\thesubsubsection.\arabic{subsubsection}}

\titleformat{\section}{\centering\normalfont\scshape}{\Roman{section}.}{5pt}{}
\titleformat{\subsection}{\normalfont\it}{\Alph{subsection}.}{5pt}{}
\titleformat{\subsubsection}[runin]{\normalfont\it}{\arabic{subsubsection})}{5pt}{}

\newcommand\infoFootnote[1]{%
  \begingroup
  \renewcommand\thefootnote{}\footnote{#1}%
  \addtocounter{footnote}{-1}%
  \endgroup}

\newtheorem{thm}{Theorem}

\newtheorem{lem}[thm]{Lemma}
\newtheorem{prop}[thm]{Proposition}

\newtheorem{defn}{Definition}

\newcommand{\N}{\mathbb{N}} 

\newcommand{\R}{\mathbb{R}}

\newcommand{\Ac}{\mathcal{A}}
\newcommand{\Bc}{\mathcal{B}}

\newcommand{\Dc}{\mathcal{D}}

\newcommand{\Fc}{\mathcal{F}} 
\newcommand{\Gc}{\mathcal{G}}

\newcommand{\Pc}{\mathcal{P}} 
\newcommand{\Rc}{\mathcal{R}} 
 
\newcommand{\Tc}{\mathcal{T}}
\newcommand{\Uc}{\mathcal{U}}

\newcommand{\Xc}{\mathcal{X}}

\newcommand{\Ab}{\boldsymbol{A}}
\newcommand{\Bb}{\boldsymbol{B}}
\newcommand{\Cb}{\boldsymbol{C}}
\newcommand{\Db}{\boldsymbol{D}}

\newcommand{\Hb}{\boldsymbol{H}}
\newcommand{\Ib}{\boldsymbol{I}}

\newcommand{\Pb}{\boldsymbol{P}}
\newcommand{\pb}{\boldsymbol{p}}
\newcommand{\Qb}{\boldsymbol{Q}}
\newcommand{\Rb}{\boldsymbol{R}}

\newcommand{\Ub}{\boldsymbol{U}}
\newcommand{\Wb}{\boldsymbol{W}}

\newcommand{\Xb}{\boldsymbol{X}}

\newcommand{\ab}{\boldsymbol{a}}
\newcommand{\bb}{\boldsymbol{b}}

\newcommand{\db}{\boldsymbol{d}}

\newcommand{\fb}{\boldsymbol{f}}
\newcommand{\gb}{\boldsymbol{g}}
\newcommand{\hb}{\boldsymbol{h}}

\newcommand{\ub}{\boldsymbol{u}}
\newcommand{\vb}{\boldsymbol{v}}

\newcommand{\xb}{\boldsymbol{x}}
\newcommand{\yb}{\boldsymbol{y}}
\newcommand{\zb}{\boldsymbol{z}}
\newcommand{\qb}{\boldsymbol{q}}

\newcommand{\oneb}{\boldsymbol{1}}
\newcommand{\zerob}{\boldsymbol{0}}

\newcommand{\xib}{\boldsymbol{\xi}}
\newcommand{\etab}{\boldsymbol{\eta}}

\newcommand{\gammab}{\boldsymbol{\gamma}}
\newcommand{\deltab}{\boldsymbol{\delta}}

\newcommand{\Phib}{\boldsymbol{\Phi}}

\newcommand{\norm}[1]{\left\lVert#1\right\rVert}

\newcommand{\interior}{\mathrm{int}}

\DeclareMathAlphabet\mathbfcal{OMS}{cmsy}{b}{n}

\title{\vspace{-2mm}\bf A mixed-integer framework for analyzing neural network-based controllers for piecewise affine systems with bounded disturbances$^\ast$}
\author{Dieter Teichrib and Moritz Schulze Darup\vspace{2mm}}
\date{}

\begin{document}

\maketitle

\textbf{\textit{Abstract}.} {\bf We present a method for representing the closed-loop dynamics of piecewise affine (PWA) systems with bounded additive disturbances and neural network-based controllers as mixed-integer (MI) linear constraints. We show that such representations enable the computation of robustly positively invariant (RPI) sets for the specified system class by solving MI linear programs. These RPI sets can subsequently be used to certify stability and constraint satisfaction. Furthermore, the approach allows to handle nonlinear systems based on suitable PWA approximations and corresponding error bounds, which can be interpreted as the bounded disturbances from above.}
\infoFootnote{D. Teichrib and M. Schulze Darup are with the \href{https://rcs.mb.tu-dortmund.de/}{Control and~Cyber-physical Systems Group}, Faculty of Mechanical Engineering, TU Dortmund University, Germany. E-mails:  \href{mailto:moritz.schulzedarup@tu-dortmund.de}{\{dieter.teichrib, moritz.schulzedarup\}@tu-dortmund.de}. \vspace{0.5mm}}
\infoFootnote{\hspace{-1.5mm}$^\ast$This paper is a \textbf{preprint} of a contribution to the 23rd European Control Conference 2025.}

\section{Introduction}

Piecewise affine (PWA) systems are an important system class in control theory as they can be used to model a large class of hybrid systems or to approximate nonlinear systems \cite{Azuma2010}. When PWA models are used to describe systems that are inherently PWA as, e.g., mixed logical dynamical (MLD) systems \cite{Bemporad2000MLDandPWA}, no error has to be considered. In these cases, analysis of the closed-loop system and control design can be performed directly using techniques developed for PWA system \cite{Mignone2000}, \cite{Grieder2005}, \cite{Camacho2010}, \cite{Teichrib2024} or MLD systems \cite{Bemporad2000MLDandPWA}. However, when a PWA model approximates a nonlinear system, an approximation error has to be considered when analysing the system behaviour and designing a controller. In \cite{Ghasemi2017}, an adaption of tube-based model predictive control (MPC) \cite{Mayne2005} is introduced, which is applicable to PWA with bounded additive disturbances. The resulting optimal control problem (OCP) is a mixed-integer program (MIP). As common in MPC, the OCP is solved in every time step, and only the first element of the optimal input sequence is applied to the system. Repeatedly solving the OCP in each time step within the sampling period of the system may be intractable for a long prediction horizon or PWA systems with many regions. Therefore, methods have been developed in \cite{Jones2009}, \cite{Chen2018}, \cite{Teichrib2024L4DC} where an approximate solution of the OCP is used instead of the exact one. Neural networks (NN) are often used in this context (see, e.g, \cite{Xiang2018,Chen2018}), as they are fast to evaluate, and there exist several software libraries (\cite{tensorflow2015-whitepaper,Pytorch}) that allow an efficient training on large data sets. However, guarantees for stability and constraint satisfaction of the original MPC do typically not hold for the approximated NN-based controller. Fortunately, in recent years, some methods to certify stability and constraint satisfaction for systems with NN-based controllers have been developed, most of them for linear systems \cite{Fazlyab2022,Schwan2023} and recently for PWA systems without disturbances \cite{Teichrib2024}. These approaches cannot be applied to PWA systems with bounded disturbances and NN-based controllers.

In this paper, we present a method based on MIP that can be used to certify stability and constraint satisfaction for PWA systems with bounded additive disturbances and NN-based controllers. Methods based on MIP are also considered in, e.g., \cite{Schwan2023} or \cite{Karg2020Reach}. However, these methods either require a stabilizing base-line controller or are only applicable to linear systems. Our method builds on the computation of polyhedral over-approximations of the $k$-step reachable set as in \cite{Teichrib2024} and, therefore, does not require a stabilizing base-line controller. These over-approximations are subsequently used to compute different robustly positively invariant (RPI) sets for certifying stability and constraint satisfaction of the closed-loop system. Furthermore, we investigate under which conditions the RPI sets for the PWA system with bounded additive disturbance allow us to show that a nonlinear system converges to a small positively invariant (PI) set while satisfying state and input constraints.

The paper is organized as follows. In the remainder of this section, we state notation and some basic definitions. In Section~\ref{sec:Fundamentals}, we give some fundamentals on mixed-integer (MI) based reachability analysis of PWA systems with NN-based controllers. The MI-based reachability analysis is extended in Section~\ref{sec:MI_PWA_with_Dis} to PWA systems with bounded additive disturbances, allowing us to handle PWA approximations of nonlinear systems. In Section~\ref{sec:AnalysisOfPWASysWithDis}, we present the main contributions of our paper, which include the computation of RPI sets for PWA systems with bounded additive disturbances and NN-based controller as well as results on stability and constraint satisfaction. Two illustrative case studies are given in Section~\ref{sec:CaseStudy}. Finally, the paper is concluded in Section~\ref{sec:Conclusions}. To improve the readability of the paper, the proofs of all lemmas are in the appendix.   

\subsection{Notation and basic definitions}
We define the support function of a polyhedron $\Xc$ for a row-vector $\vb\in\R^{1\times n}$ as $h_{\Xc}(\vb):=\sup_{\xb\in\Xc} \vb\xb$. For a matrix argument $\Hb\in\R^{w\times n}$, $h_{\Xc}(\Hb)$ is understood as $h_{\Xc}(\Hb)=\begin{pmatrix} h_{\Xc}(\Hb_1) & \dots & h_{\Xc}(\Hb_w) \end{pmatrix}^\top$, where $\Hb_i$ refers to the $i$-th row of $\Hb$. The scaling of a set $\Ac$ by a scalar $s$ is defined as $s \Ac:=\{s \xb \in \R^n \ | \ \xb \in \Ac \}$. For compact and convex sets, the operation $\Ac\oplus\Bc:=\{ \ab+\bb \ | \ \ab\in\Ac, \ \bb\in\Bc \}$ is the Minkowski sum. Natural numbers and natural numbers, including $0$, are denoted by $\N$ and $\N_0$, respectively. The operator $\text{col}(\xb(j)_{j=K_1}^{K_2})$ stacks the column vectors $\xb(j)$ with $j\in\{K_1,\dots,K_2\}$ in a single column vector. 
We further define $d(\zb,\Xc):=\inf_{\xb\in\Xc} \norm{\zb-\xb}_2$ as the distance of a point to a set and $\Bc_\delta:=\{ \xb\in\R^n \ | \ d(\xb,\zerob)\leq \delta \}$ as a $2$-norm ball with radius $\delta>0$. Moreover, we sometimes use the short-hand notation $\zerob$ for a set containing only the origin. 

\section{Fundamentals on mixed-integer based reachability analysis}\label{sec:Fundamentals}

In this paper, we consider discrete-time PWA systems
\begin{equation}
    \label{eq:SysWithDis}
    \xb(k+1)=\fb_{\text{PWA}}(\xb(k),\ub(k)) + \db(k)
\end{equation}
with 
\begin{equation*}
    \fb_{\text{PWA}}:=\Ab^{(i)}\xb(k)+\Bb^{(i)}\ub(k)+\pb^{(i)} \,\,\,\text{if} \ \begin{pmatrix}
        \xb(k) \\
        \ub(k)
    \end{pmatrix}\!\in\!\Pc^{(i)}.
\end{equation*}
with $i\in\{1,\dots,s\}$. The additive disturbance $\db(k)$ is assumed to be bounded by state and input-dependent polyhedral sets, i.e., $\db(k)\in\Dc^{(i)}\subset\R^n$. The polyhedral sets 
\begin{equation}\label{eq:HRep}
\Pc^{(i)}:=\{ \xib \in \R^{n+m} \ | \ \Hb^{(i)} \xib \leq \hb^{(i)} \}
\end{equation}
with $\Hb^{(i)}\in\R^{d_i\times (n+m)}$ and $\hb^{(i)}\in\R^{d_i}$ for all $i\in\{1,\dots,s\}$
are pairwise disjoint, i.e, $\interior(\Pc^{(i)})\cap\interior(\Pc^{(j)})=\emptyset$ for all $i\neq j$, and partition the state $\Xc\subset\R^n$ and input space $\Uc\subset\R^m$ according to $\Xc \times \Uc=\cup_{i=1}^s\Pc^{(i)}$ where $s$ is the number of polyhedral sets. We aim to develop methods for computing large, robustly positively invariant (RPI) sets contained in $\Xc$, in which the controlled system \eqref{eq:SysWithDis} with an NN-based controller
\begin{equation}\label{eq:UOfK}
    \ub(k)=\Phib(\xb(k))
\end{equation}
can be operated safely for all time. Where safe means that the state and input constraints are satisfied. Moreover, we aim to compute small RPI sets to which the controlled system converges. The methods mainly build on the $k$-step reachable set 
\begin{align}
    \nonumber
    \Rc^\Dc_k(\Fc):=\{\xb^+\in \ \R^n \ |& \ \xb^+=\fb_{\text{PWA}}(\xb,\Phib(\xb))+\db,\\
    \label{eq:kStepReachSet}
    & \ \xb\in \Rc^{\Dc}_{k-1}(\Fc),\db\in\Dc^{(i)}\}
\end{align}
with $\Rc^\Dc_0(\Fc):=\Fc$ or over-approximations of this set. Based on the $k$-step reachable set \eqref{eq:kStepReachSet}, we can define an RPI set as follows. 
\begin{defn}[{\cite[Def.~1]{rakovic2005invariant}}]\label{def:RPI}
    A Set $\Fc\subseteq\R^n$ is RPI if and only if 
    \begin{equation}\label{eq:RPI}
        \Rc^\Dc_1(\Fc)\subseteq\Fc.
    \end{equation}
\end{defn}

To compute the reachable sets \eqref{eq:kStepReachSet} via mixed-integer programming (MIP), we will summarize in the following subsections MI linear constraints that allow a description of NN-based controllers \eqref{eq:UOfK} and PWA systems without disturbances, i.e., $\db(k)=\zerob$. This description is extended to the case $\db(k)\in\Dc^{(i)}$ in Section \ref{sec:MI_PWA_with_Dis}.

\subsection{MI formulation of neural networks}

We consider feed-forward-NN of the form 
\begin{equation}\label{eq:NN}
    \Phib(\xb):=\fb^{(\ell+1)}\circ \gb^{(\ell)}\circ \fb^{(\ell)}\circ \dots \circ \gb^{(1)}\circ \fb^{(1)}(\xb)
\end{equation}
where $\ell$ is the number of hidden layers and $w_i$ the number of neurons in layer $i$. Here, the functions $\fb^{(i)}: \R^{w_{i-1}} \rightarrow \R^{p_i w_i}$ for $i \in \{1,\dots,\ell\}$ refer to preactivations, where the parameter $p_i\in \N$ allows to consider ``multi-channel'' preactivations as required for maxout activation (see \cite{Goodfellow2013}). Moreover, $\gb^{(i)}: \R^{p_i w_i} \rightarrow \R^{w_i}$ stand for activation functions and $\fb^{(\ell+1)}: \R^{w_{\ell}} \rightarrow \R^{w_{\ell+1}}$ reflects postactivation.
We assume affine functions $\fb^{(i)}$, i.e., 
\begin{equation}\label{eq:eq:PWAPreactivation}
    \fb^{(i)}(\yb^{(i-1)}):=\Wb^{(i)}\yb^{(i-1)}+\bb^{(i)},
\end{equation}
where $\Wb^{(i)}\!\!\in\R^{p_i w_i\times w_{i-1}}\!$ is a weighting matrix, ${\bb^{(i)} \!\!\in \R^{p_i w_i}}$ is a bias vector, and $\yb^{(i-1)}$ denotes the output of the previous layer with $\yb^{(0)}:=\xb \in \R^n$. Various choices for the activation functions have been proposed. However, since we aim for a mixed-integer formulation of the NN, we focus on PWA activation functions. More precisely, we consider the maxout activation function 
\begin{equation}
    \label{eq:maxout}
    \gb^{(i)}(\zb^{(i)}):=
    \begin{pmatrix}
        \max \limits_{1\leq j \leq p_i}\big\{\zb_j^{(i)}\big\} \\
        \vdots \\
        \max \limits_{p_i(w_i-1)+1 \leq j \leq p_i w_i}\big\{\zb_{j}^{(i)}\big\}
    \end{pmatrix},
\end{equation} 
since the maxout activation is more general than other PWA activation functions and includes, e.g., ReLU or leaky ReLU, as special cases. We refer to the resulting NN as a maxout NN. We assume that $\Phib(\xb)\in\Uc$ holds for all $\xb\in\Xc$. Note that for polyhedral state constraints as considered here, a given NN can always be modified to satisfy the required condition by adding layers to the NN that perform a projection onto the input constraints. These layers can be constructed using methods from, e.g., \cite[Sec.~III-B]{Markolf2021} or \cite[Sec.~II-B]{Teichrib2024}. For analyzing a PWA system with an NN-based controller, we need a description of the maxout NN in terms of MI linear constraints. According to \cite[Lem.~2]{Teichrib2023Ifac} the MI linear constraints
\begin{subequations}\label{eq:MI_Constraints_NN}
\begin{align}
    \qb^{(i)}_l(k)\! &\leq\! \Wb^{(i)}_j \qb^{(i-1)}(k)\! +\! \bb^{(i)}_j\! +\! \overline{b}^{(i)} (1\!-\!\deltab^{(i)}_j(k)), \\
    -\qb^{(i)}_l(k) &\leq -\Wb^{(i)}_j \qb^{(i-1)}(k)\! -\! \bb^{(i)}_j, \\
    \sum\limits_{\jmath\in\Ac^{(i)}_l} &\deltab^{(i)}_\jmath(k) =1, \quad  \\
    \forall j &\in \Ac^{(i)}_l, \ \forall l \in \{1,\dots,w_i\},  \forall i \in \{1,\dots,\ell\},
\end{align} 
\end{subequations}
with $\Ac^{(i)}_l:=\{p_i(l-1)+1,\dots,p_i l\}$ and $\deltab^{(i)}(k)\in\{0,1\}^{p_i w_i}$ are such that the output of the NN is $\Phib(\qb^{(0)}(k))=\Wb^{(\ell+1)}\qb^{(\ell)}(k)+\bb^{(\ell+1)}$.
Thus, the desired MI linear constraints are given by \eqref{eq:MI_Constraints_NN}.

\subsection{MI formulation of PWA systems without disturbances}

In this section, we briefly summarize known methods from \cite{Bemporad1999} for constructing MI linear constraints for PWA systems of the form \eqref{eq:SysWithDis} with $\db(k)=\zerob$. The first step is to introduce the MI linear constraints
\begin{subequations}\label{eq:gammaForR}
\begin{align}
    \Hb^{(i)} \begin{pmatrix} \xb(k) \\ \ub(k) \end{pmatrix} &\leq \hb^{(i)} + \oneb M (1-\gammab_i(k)) \\
    \oneb^\top \gammab(k) &= 1, \quad \gammab(k)\in\{0,1\}^s
\end{align}
\end{subequations}
for all $i\in\{1,\dots,s\}$ and all $k\in\{0,\dots,K-1\}$ with $K\in\N$ and $\gammab(k)\in\{0,1\}^s$, which are according to \cite{Bemporad1999} such that 
\begin{equation}\label{eq:GammaXinP}
\gammab_i(k)=1\Leftrightarrow\begin{pmatrix} \xb^\top(k) & \ub^\top(k) \end{pmatrix}^\top \in \Pc^{(i)}
\end{equation}
holds. The constant $M$ in the constraints \eqref{eq:gammaForR} is often referred to as big-$M$ and can be chosen according to \cite[Eq.~(20)-(21)]{Bemporad1999}. Now, the binary variable $\gammab(k)$ can be used to describe the PWA system dynamics \eqref{eq:SysWithDis} with $\db(k)=\zerob$ by the MI linear constraints
\begin{subequations}\label{eq:MISys}
\begin{align}
    \nonumber
    \hspace{-4mm}-\oneb M(1\!-\!\gammab_i(k))\!&\leq\!\Ab^{(i)}\xb(k)\!+\!\Bb^{(i)}\ub(k)\!+\!\pb^{(i)}\!-\!\tilde{\xb}^{(i)}(k\!+\!1) \\
    \label{eq:MI_PWA_ABp}
    &\leq \oneb M(1-\gammab_i(k)) \\
    \label{eq:MI_PWA_xTilde}
    - \oneb M \gammab_i(k) &\leq \tilde{\xb}^{(i)}(k)\leq \oneb M \gammab_i(k) \\
    \label{eq:MI_PWA_sumXTilde}
    \xb(k+1) &= \sum\limits_{j=1}^s \tilde{\xb}^{(j)}(k+1) 
\end{align}
\end{subequations}
for all $i\in\{1,\dots,s\}$ and all $k\in\{0,\dots,K-1\}$ (cf. \cite[Sec.~3.1]{Bemporad1999}), which are such that $\xb(k+1)=\fb_{\text{PWA}}(\xb(k),\ub(k))$ holds for all $k\in\{0,\dots,K-1\}$.

\section{Mixed-integer linear constraints for PWA systems with bounded disturbances}\label{sec:MI_PWA_with_Dis}

For computing reachable sets for \eqref{eq:SysWithDis} by mixed-integer linear programs (MILP), we extend the MI linear constraints \eqref{eq:MISys} to systems with state and input-dependent disturbances of the form   
\begin{equation}\label{eq:dinDi}
    \db(k) \in \Dc^{(i)} \ \text{if} \ \begin{pmatrix}
        \xb(k)^\top &
        \ub(k)^\top
    \end{pmatrix}^\top \in \Pc^{(i)}.
\end{equation}
This type of systems can be described by the MI linear constraints \eqref{eq:gammaForR}, \eqref{eq:MI_PWA_xTilde}, \eqref{eq:MI_PWA_sumXTilde}, and 
\begin{subequations}\label{eq:MISysWithDis}
\begin{align}
    \nonumber
    -\oneb M(1\!-\!\gammab_i(k))\leq&\Ab^{(i)}\xb(k)+\Bb^{(i)}\ub(k)+\pb^{(i)}+\db(k)\\
    \label{eq:MI_PWA_ABpd}
    &-\tilde{\xb}^{(i)}(k+1)\leq \oneb M(1\!-\!\gammab_i(k)) \\
    \label{eq:MI_PWA_Hd}
    \Hb^{(\Dc^{(i)})} \db(k) \leq& \hb^{(\Dc^{(i)})} + \oneb M (1-\gammab_i(k)).
\end{align}
\end{subequations}
\begin{lem}\label{lem:MI_PWA_with_dis}
    Consider the system dynamics \eqref{eq:SysWithDis} with \eqref{eq:dinDi}, a maxout NN $\Phib(\xb)$ as in \eqref{eq:NN}--\eqref{eq:maxout}, and define $\Xb_{K+1}=\text{col}(\xb(j)_{j=0}^{K})$, $\Ub_{K}=\text{col}(\ub(j)_{j=0}^{K-1})$, $\Db_{K}=\text{col}(\db(j)_{j=0}^{K-1})$, and $\tilde{\Xb}_{K}^{(i)}=\text{col}(\tilde{\xb}^{(i)}(j)_{j=1}^{K})$ 
    for all $i\in\{1,\dots,s\}$. Then, any solution to the MI feasibility problem 
    \begin{subequations}\label{eq:MIClosedLoop}
    \begin{align}
        \text{find} \ &\!\Xb_{K+1},\!\Ub_{K},\!\Db_{K},\!\tilde{\Xb}_{K}^{(1)}\!,\!...,\!\tilde{\Xb}_{K}^{(s)}\!,\!\gammab(0),\!...,\!\gammab(K\!-\!1)\\
        \nonumber
        &\qb^{(1)}(0),\dots,\qb^{(\ell)}(0),\dots,\qb^{(1)}(K-1),\dots,\qb^{(\ell)}(K-1)\\
        \nonumber
        &\deltab^{(1)}(0),\dots,\deltab^{(\ell)}(0),\dots,\deltab^{(1)}(K-1),\dots,\deltab^{(\ell)}(K-1) 
    \end{align}
    \begin{align}
        \label{eq:MIConClosedLoop1}
        \text{s.t.} \ &\text{\eqref{eq:MI_Constraints_NN},} \text{ \eqref{eq:gammaForR},} \text{ \eqref{eq:MI_PWA_xTilde},} \text{ \eqref{eq:MI_PWA_sumXTilde},} \text{ \eqref{eq:MISysWithDis},} \hspace{23mm} \\
        \label{eq:MIConQEqhatX}
        \qb^{(0)}(k)&= \xb(k),\\ 
        \label{eq:hatUEqWqplusB}
        \ub(k)&=\Wb^{(\ell+1)}\qb^{(\ell)}(k)+\bb^{(\ell+1)}
    \end{align}
    \end{subequations}
    for all $k\in\{0,\dots,K-1\}$, is such that
    \begin{subequations}\label{eq:PWAConstraints}
    \begin{align}
        \label{eq:CLSystem}
        \xb(k+1)&=\fb_{\text{PWA}}(\xb(k),\Phib(\xb(k)))+\db(k), \\
        \label{eq:dInDLemma}
        \db(k)&\in\Dc^{(i)},
    \end{align}
    \end{subequations}
    hold for all $k\in\{0,\dots,K-1\}$ and for all $\xb(0)\in\R^n$.
\end{lem}

For OP where the PWA system dynamics \eqref{eq:CLSystem} with bounded disturbance \eqref{eq:dInDLemma} is in the constraints, Lemma~\ref{lem:MI_PWA_with_dis} can be used to replace the nonlinear constraints \eqref{eq:CLSystem}-\eqref{eq:dInDLemma} by the MI linear constraints \eqref{eq:MIConClosedLoop1}-\eqref{eq:hatUEqWqplusB}, thus transforming the nonlinear OP to an MILP that can be solved with standard software, e.g., \cite{mosek}. This can be used to formulate a MILP for the evaluation of the support function of the $k$-step reachable set for the PWA system \eqref{eq:SysWithDis} with \eqref{eq:UOfK}, i.e. a PWA system with bounded disturbance and an NN-based controller. 
\begin{lem}\label{lem:suppF_Rk}
    Let the $k$-step reachable set be defined as in \eqref{eq:kStepReachSet}. Then the MILP
    \begin{subequations}\label{eq:MILPReachableSets}
    \begin{align}
        c_k^\ast(\vb,\Xc):=&\hspace{-5mm}\max_{\substack{\Xb_{k+1},\Ub_{k},\Db_{k},\tilde{\Xb}_{k}^{(1)},\dots,\tilde{\Xb}_{k}^{(s)},\gammab(0),\dots,\gammab(k-1)\\ \qb^{(1)}(0),\dots,\qb^{(\ell)}(0),\dots,\qb^{(1)}(k-1),\dots,\qb^{(\ell)}(k-1)\\
        \deltab^{(1)}(0),\dots,\deltab^{(\ell)}(0),\dots,\deltab^{(1)}(k-1),\dots,\deltab^{(\ell)}(k-1)}} \hspace{-5mm}\vb \xb(k) \\
        &\text{s.t.} \quad \text{\eqref{eq:MIConClosedLoop1}--\eqref{eq:hatUEqWqplusB}}
    \end{align}
    \end{subequations}
    with $\vb\in\R^{1\times n}$, is such that $c_k^\ast(\vb,\Xc)=\hb_{\Rc^\Dc_k(\Xc)}(\vb)$ holds.
\end{lem}

Based on the support function $\hb_{\Rc^\Dc_1(\Xc)}$ of the $1$-step reachable set $\Rc^\Dc_1(\Xc)$, we define an over-approximation
\begin{equation}\label{eq:Rover1}
    \overline{\Rc}^\Dc_1(\Xc):=\{\xb\in\R^n \ | \ \Cb\xb\leq \hb_{\Rc^\Dc_1(\Xc)}(\Cb) \}
\end{equation}
with $\Cb\in\R^{l\times n}$, which is according to \cite[Lem.~3]{Teichrib2024} such that $\Rc^\Dc_1(\Xc)\subseteq\overline{\Rc}^\Dc_1(\Xc)$ and $\hb_{\Rc^\Dc_{1}(\Xc)}(\Cb)=\hb_{\overline{\Rc}^\Dc_{1}(\Xc)}(\Cb)$ hold. The hyperplanes in the matrix $\Cb$ are a design parameter. A typical choice are hyperplanes that lead to a hypercube, i.e., $\Cb=(\Ib \ -\Ib)^\top$. By using Lemma~\ref{lem:suppF_Rk}, the set \eqref{eq:Rover1} can be computed by solving $l$-times the MILP~\eqref{eq:MILPReachableSets}. Moreover, with Lemma~\ref{lem:suppF_Rk}, it is possible to check whether a given polyhedral set $\Fc$ is RPI. According to Definition~\ref{def:RPI}, checking if a given set is RPI requires verifying the condition $\Rc_1^\Dc(\Fc)\subseteq\Fc$. For a polyhedral set $\Fc$ with $l$ hyperplanes, \eqref{eq:RPI} holds if and only if $\hb_{\Rc^\Dc_{1}(\Fc)}(\Hb^{(\Fc)}) \leq \hb^{(\Fc)}$, which can be verified by solving $l$-times the MILP~\eqref{eq:MILPReachableSets}.   

\section{Analysis of PWA systems with bounded disturbances and NN-based controllers}\label{sec:AnalysisOfPWASysWithDis}
The analysis in this section is based on the over-approximation 
\begin{equation}\label{eq:ROverOver}
    \overline{\Rc}_k^\Dc(\Fc):=\underbrace{\overline{\Rc}^\Dc_1(\overline{\Rc}^\Dc_1(\dots\overline{\Rc}^\Dc_1}_{k \ \text{times}}(\Fc))) 
\end{equation}
with $\overline{\Rc}_0^\Dc(\Fc):=\Fc$ of the $k$-step reachable set \eqref{eq:kStepReachSet}. According to \cite[Lem.~5]{Teichrib2024} and \cite[Cor.~4]{Teichrib2024} the set $\overline{\Rc}_k^\Dc(\Fc)$ is such that 
\begin{subequations}\label{eq:RoveroverFeat}
\begin{align}
\label{eq:RkSubsetEqRkOver}
\Rc_k^\Dc(\Ac)&\subseteq\overline{\Rc}_k^\Dc(\Ac) \ \text{for all} \  k\in\N_0 \ \text{and} \\ \overline{\Rc}_1(\Ac)&\subseteq\overline{\Rc}_1(\Bc) \ \text{for} \ \Ac\subseteq\Bc. 
\end{align}
\end{subequations}
These two features are crucial for the following results and are often used in the proofs. The over-approximation \eqref{eq:ROverOver} builds on the repeated computation of the one-step reachable set. Thus, evaluating the support function of $\overline{\Rc}_{\hat{k}}^\Dc(\Fc)$ requires to solve the MILP~\eqref{eq:MILPReachableSets} with $k=1$ and $N_1:=s+\sum_{i=1}^{\ell}p_iw_i$ binary variables $\hat{k}$-times. Whereas evaluating the support function of $\Rc_{\hat{k}}^\Dc(\Fc)$ requires solving a larger MILP~\eqref{eq:MILPReachableSets} with $k=\hat{k}$ and thus $\hat{k}N_1$ binary variables only once. Meaning the computational complexity of evaluating the support function of $\overline{\Rc}_{\hat{k}}^\Dc(\Fc)$ grows linearly with $\hat{k}$, whereas for $\Rc_{\hat{k}}^\Dc(\Fc)$ it grows in the worst case exponentially with $\hat{k}$, due to the increasing number of binary variables. Thus, for computational reasons, all results are formulated for the over-approximation \eqref{eq:ROverOver} of the $k$-step reachable set \eqref{eq:kStepReachSet}. However, since the following proofs apply to sets satisfying the conditions \eqref{eq:RoveroverFeat}, the results still hold if $\overline{\Rc}_{k}^\Dc(\Fc)$ is replaced by the exact $k$-step reachable set $\Rc_{k}^\Dc(\Fc)$.

\subsection{Computation of a safe set}\label{sec:SafeSet}
We start with computing a large RPI set in which the closed-loop system can be operated safely, i.e., without violating any constraints, for all time. The largest set with the required specifications is the maximum RPI set. However, this set cannot be computed using the techniques introduced in Section~\ref{sec:MI_PWA_with_Dis}. Therefore, inspired by \cite[Alg.~1]{Teichrib2024} we define the following set
\begin{equation}\label{eq:Fk}
    \Fc_{k+1}=\overline{\Rc}^\Dc_1(\Fc_k)\cap\Xc \ \text{with} \ \Fc_0:=\Xc.
\end{equation}
\begin{prop}\label{prop:Fmax}
    Let $\Fc_k$ be as in \eqref{eq:Fk} and assume that there exists a $k^\ast$ such that $\overline{\Rc}^\Dc_{k^\ast+1}(\Xc)\subseteq\Xc$. Then, there exist a $\hat{k}\leq k^\ast$ such that $\Fc_{\text{max}}:=\Fc_{\hat{k}}$ is a RPI set with $\Fc_{\text{max}}\subseteq\Xc$. 
\end{prop}
\begin{proof}
    We first note that $\Fc_k\subseteq\overline{\Rc}_k^\Dc(\Xc)$ and $\Fc_{k+1}\subseteq\Fc_k\subseteq\Xc$ for all $k\in\N_0$. Thus we have
    \begin{equation}\label{eq:R1FkAstinX}
        \overline{\Rc}_1^\Dc(\Fc_{k^\ast})\subseteq\overline{\Rc}_1^\Dc(\overline{\Rc}_{k^\ast}^\Dc(\Xc))=\overline{\Rc}_{k^\ast+1}^\Dc(\Xc)\subseteq\Xc
    \end{equation}
    which implies
    \begin{equation}\label{eq:FkAstRPI}
        \Rc_1^\Dc(\Fc_{k^\ast})\!\subseteq\!\overline{\Rc}_1^\Dc(\Fc_{k^\ast})\!=\!\overline{\Rc}_1^\Dc(\Fc_{k^\ast})\cap\Xc\!=\!\Fc_{k^\ast+1}\!\subseteq\!\Fc_{k^\ast}.
    \end{equation}
    As apparent from \eqref{eq:R1FkAstinX} and \eqref{eq:FkAstRPI} $\Fc_{k^\ast}$ is RPI if $\overline{\Rc}_1^\Dc(\Fc_{k^\ast})\subseteq\Xc$ holds. Thus the condition $\overline{\Rc}^\Dc_{k^\ast+1}(\Xc)\subseteq\Xc$ is only sufficient for $\Fc_{k^\ast}$ to be RPI, i.e., there may exist a $\hat{k}\leq k^\ast$ for which $\Fc_{\hat{k}}$ is RPI.  
\end{proof}
According to Proposition~\ref{prop:Fmax}, a large RPI set can be computed by iterating \eqref{eq:Fk} until $\overline{\Rc}_1^\Dc(\Fc_k)\subseteq\Fc_k$ holds and thus $\Fc_k$ is RPI. This is the case after, at most, $k^\ast$ iterations. 

\subsection{Computation of a terminal set}\label{sec:TerminalSet}
The next step is to compute a small terminal set, which is RPI and to which all trajectories starting in $\Fc_{\text{max}}$ converge. For stable linear systems, a set with the required specifications is given by the minimal RPI set $\Rc_{\text{min}}:=\overline{\Rc}_{\infty}^\Dc(\zerob)$, which is typically not finitely determined. Thus, often RPI over-approximations $\underline{\Rc}_{\text{min}}:=(1+\underline{\epsilon})\overline{\Rc}_{\underline{k}}^\Dc(\zerob)$ with $\underline{\epsilon}>0$ are considered, where $\underline{k}$ is chosen depending on $\underline{\epsilon}$ such that $\underline{\Rc}_{\text{min}}$ is RPI and an over-approximation of the minimal RPI set (cf. \cite{rakovic2005invariant}). However, the set $\underline{\Rc}_\text{min}$ can only be used as terminal set if there exists a $k$ with $\Rc_k^\Dc(\Fc_{\text{max}})\subseteq\underline{\Rc}_\text{min}$, which is not guaranteed for PWA systems. Thus, there may exist $\underline{\Rc}_\text{min}$ that are not reachable for the closed-loop system from $\Fc_{\text{max}}$. Therefore, we alternatively consider the set 
\begin{equation}\label{eq:RminOver}
    \overline{\Rc}_\text{min}:=\overline{\Rc}_{\overline{k}}^\Dc(\Fc_\text{max}) 
\end{equation}
where $\overline{k}$ is chosen so that 
\begin{equation}\label{eq:TerminationFmin}
    \frac{1}{1+\overline{\epsilon}}\overline{\Rc}_{\overline{k}}^\Dc(\Fc_{\text{max}})\subseteq\overline{\Rc}_1^\Dc\left(\frac{1}{1+\overline{\epsilon}}\overline{\Rc}_{\overline{k}}^\Dc(\Fc_{\text{max}})\right)
\end{equation}
holds for a given $\overline{\epsilon}>0$. Then, according to \cite[Prop.~7]{Teichrib2024}, $\overline{\Rc}_\text{min}$ is an RPI over-approximation of $\overline{\Rc}_{\infty}^\Dc(\Fc_{\text{max}})$ with
\begin{equation*}
    \overline{\Rc}_{\infty}^\Dc(\Fc_{\text{max}})\subseteq\overline{\Rc}_\text{min}\subseteq(1+\overline{\epsilon})\overline{\Rc}_{\infty}^\Dc(\Fc_{\text{max}}).
\end{equation*}
Moreover, since $\xb(k)\in\Rc_{\overline{k}}^\Dc(\Fc_\text{max})\subseteq\overline{\Rc}_{\overline{k}}^\Dc(\Fc_\text{max})=\overline{\Rc}_\text{min}$ holds for all $\xb(0)\in\Fc_\text{max}$ and $k\geq\overline{k}$, all trajectories starting in $\Fc_\text{max}$ will enter and remain in $\overline{\Rc}_\text{min}$ after at most $\overline{k}$ time steps. Thus the set $\overline{\Rc}_\text{min}$ is reachable for the closed-loop system for all $\xb(0)\in\Fc_\text{max}$. 

\subsection{Stability of the closed-loop system}
Now, we combine the previous results to state one of the main results of this section that proves stability of the set $\overline{\Rc}_{\text{min}}$ for the PWA system with bounded disturbance \eqref{eq:SysWithDis} and NN-based controller \eqref{eq:UOfK}. Where stability of a set is defined based on \cite[Sec.~1]{Mayne2005} as follows: A set $\Tc$ is (Lyapunov) stable if for every $\epsilon>0$, there exists a $\delta>0$ such that, if $d(\xb(0),\Tc)<\delta$, then $d(\xb(k),\Tc)<\epsilon$ for all $k\in\N_0$. If $\Tc$ is stable and $\lim\limits_{k\to\infty}d(\xb(k),\Tc)=0$ holds for all $\xb(0)\in\Fc$, then $\Tc$ is asymptotically stable with region of attraction $\Fc$.
\begin{thm}\label{thm:BoundedStab}
    Let $\Fc_{\text{max}}$ and $\overline{\Rc}_{\text{min}}$ by defined as in Proposition \ref{prop:Fmax} and \eqref{eq:RminOver}, respectively and assume that 
    \begin{equation}\label{eq:StabCondFmin}
        \alpha \overline{\Rc}_{\overline{k}}^\Dc(\Fc_{\text{max}}) \subseteq \overline{\Rc}_{\overline{k}-1}^\Dc(\Fc_{\text{max}})
    \end{equation}
    holds for $\alpha>1$. Then, the set $\overline{\Rc}_{\text{min}}$ is asymptotically stable for the system \eqref{eq:SysWithDis} with $\ub(k)=\Phib(\xb(k))$. The region of attraction is $\Fc_{\text{max}}$. Moreover, $\xb(k)\in\Xc$ and $\ub(k)\in\Uc$ holds for all $k\in \N_0$ with $\xb(0)\in\Fc_{\text{max}}$.
\end{thm}
\begin{proof}
    Since $\Fc_{\text{max}}\subseteq\Xc$ is an RPI set for the closed-loop system \eqref{eq:CLSystem}, we have for all $\xb(0)\in\Fc_{\text{max}}$ that $\xb(k)\in\Fc_{\text{max}}\subseteq\Xc$ holds for all $k\in\N$. Moreover, by assumption we then have $\Phib(\xb)\in\Uc$ for all $\xb\in\Xc$ and thus $\Phib(\xb(k))\in\Uc$ for all $k\in\N_0$. To prove asymptotic stability, we first define $\delta_{\text{max}}:=\max \delta$ subject to $\overline{\Rc}^\Dc_{\overline{k}}(\Fc_{\text{max}}) \oplus \Bc_\delta \subseteq \alpha \overline{\Rc}^\Dc_{\overline{k}}(\Fc_{\text{max}})$ and choose $\delta=\min\{ \epsilon,\delta_{\text{max}} \}$ with $\epsilon>0$. Note that $\delta_{\text{max}}$ can be explicitly computed using the support function and is given by $\delta_{\text{max}}=(\alpha-1)\min_{\etab} h_{\overline{\Rc}^\Dc_{\overline{k}}(\Fc_{\text{max}})}(\etab)$. For polyhedral $\overline{\Rc}^\Dc_{\overline{k}}(\Fc_{\text{max}})$ with hyperplanes $\Cb$, $\delta_{\text{max}}$ simplifies to $\delta_{\text{max}}=(\alpha-1)\min_{i} h_{\overline{\Rc}^\Dc_{\overline{k}}(\Fc_{\text{max}})}(\Cb_i)$. Since $\alpha>1$ and $\overline{\Rc}^\Dc_{\overline{k}}(\Fc_{\text{max}})$ has, due to $\overline{\Rc}^\Dc_{\overline{k}}(\zerob)\subseteq\overline{\Rc}^\Dc_{\overline{k}}(\Fc_{\text{max}})$, a nonempty interior, we have $\delta_{\text{max}}>0$. We further have 
    \begin{align}
    \nonumber
        \overline{\Rc}^\Dc_k\big(\overline{\Rc}^\Dc_{\overline{k}}(\Fc_{\text{max}}) \oplus \Bc_\delta\big) &\subset \overline{\Rc}^\Dc_{\overline{k}}(\Fc_{\text{max}}) \oplus \Bc_\delta\\
        \label{eq:PlusDeltaPosInv}
        &\subseteq \overline{\Rc}^\Dc_{\overline{k}}(\Fc_{\text{max}}) \oplus \Bc_\epsilon
    \end{align}
    for all $k\in\N$, for all $\epsilon>0$ and $\delta=\min\{ \epsilon,\delta_{\text{max}} \}$. For $k=1$, \eqref{eq:PlusDeltaPosInv} holds due to   
    \begin{align*}
        &\overline{\Rc}^\Dc_1\big(\overline{\Rc}^\Dc_{\overline{k}}(\Fc_{\text{max}}) \oplus \Bc_\delta\big) \subseteq\overline{\Rc}^\Dc_1(\alpha\overline{\Rc}^\Dc_{\overline{k}}(\Fc_{\text{max}}))\subseteq\overline{\Rc}^\Dc_{\overline{k}}(\Fc_{\text{max}})\\
        \subset&\overline{\Rc}^\Dc_{\overline{k}}(\Fc_{\text{max}}) \oplus \Bc_\delta \subseteq \overline{\Rc}^\Dc_{\overline{k}}(\Fc_{\text{max}}) \oplus \Bc_\epsilon
    \end{align*}
    for all $\epsilon>0$ and $\delta=\min\{ \epsilon,\delta_{\text{max}} \}\leq\epsilon$. Thus, \eqref{eq:PlusDeltaPosInv} holds for one $i$. For $k=i+1$ we have 
    \begin{align*}
        &\overline{\Rc}^\Dc_1\Big(\overline{\Rc}^\Dc_{i}\big(\overline{\Rc}^\Dc_{\overline{k}}(\Fc_{\text{max}}) \oplus \Bc_\delta\big)\Big) \subseteq \overline{\Rc}^\Dc_1\big(\overline{\Rc}^\Dc_{\overline{k}}(\Fc_{\text{max}}) \oplus \Bc_\delta\big)\\
        \subset&\overline{\Rc}^\Dc_{\overline{k}}(\Fc_{\text{max}}) \oplus \Bc_\delta \subseteq \overline{\Rc}^\Dc_{\overline{k}}(\Fc_{\text{max}}) \oplus \Bc_\epsilon
    \end{align*}
    for all $\epsilon>0$. As a result, \eqref{eq:PlusDeltaPosInv} holds for all $k\in\N$. Now, if $d(\xb(0),\overline{\Rc}_{\text{min}})<\delta$, then $\xb(0)\in\overline{\Rc}^\Dc_{\overline{k}}(\Fc_{\text{max}})\oplus \Bc_\delta$. Consequently, according to \eqref{eq:PlusDeltaPosInv} we have $\xb(k) \in \Rc^\Dc_k\big(\overline{\Rc}^\Dc_{\overline{k}}(\Fc_{\text{max}}) \oplus \Bc_\delta\big) \subseteq \overline{\Rc}^\Dc_k\big(\overline{\Rc}^\Dc_{\overline{k}}(\Fc_{\text{max}}) \oplus \Bc_\delta\big) \subset \overline{\Rc}^\Dc_{\overline{k}}(\Fc_{\text{max}}) \oplus \Bc_\epsilon$ and thus $d(\xb(k),\overline{\Rc}_{\text{min}})<\epsilon$ for all $k\in\N$ and for all $\epsilon>0$. For $k=0$ we directly obtain $d(\xb(0),\overline{\Rc}_{\text{min}})<\delta\leq\epsilon$. Thus, in summary for every $\epsilon>0$ there exist a $\delta=\min\{\epsilon,\delta_{\text{max}}\}$ such that if $d(\xb(0),\overline{\Rc}_{\text{min}}) < \delta$, then $d(\xb(k),\overline{\Rc}_{\text{min}}) < \epsilon$ for all $k\in\N_0$. This proves stability of the system. Since $\Rc^\Dc_{\overline{k}}(\Fc_{\text{max}})\subseteq\overline{\Rc}^\Dc_{\overline{k}}(\Fc_{\text{max}})=\overline{\Rc}_{\text{min}}$ hold, we further have $\xb(k)\in \overline{\Rc}_{\text{min}}$ for all $\xb(0)\in\Fc_{\text{max}}$ and for all $k\geq\overline{k}$. Therefore $\lim\limits_{k\to\infty}d(\xb(k),\overline{\Rc}_{\text{min}})=0$ for all $\xb(0)\in\Fc_{\text{max}}$ and thus $\overline{\Rc}_{\text{min}}$ is asymptotic stable with region of attraction $\Fc_{\text{max}}$. 
\end{proof}

\subsection{Analysis of nonlinear systems}\label{sec:AnalysisOfNLSys}

The proposed approach for reachability analysis can be used to analyze nonlinear systems that can be approximated by PWA systems with bounded additive disturbances. Therefore, we define the $k$-step reachable set of the nonlinear system 
\begin{equation}\label{eq:f}
\xb(k+1)=\fb(\xb(k),\ub(k))
\end{equation}
with an NN-based controller as in \eqref{eq:UOfK} as follows
\begin{equation*}
    \Rc_k(\Fc)\!:=\!\{\xb^+\!\in\!\R^n \ \!|\! \ \xb^+\!=\!\fb(\xb,\Phib(\xb)),\xb\!\in\!\Rc_{k-1}(\Fc)\}.
\end{equation*}
with $\Rc_0(\Fc)=\Fc$ and state a result that relates the reachable sets of \eqref{eq:f} and the reachable sets of \eqref{eq:SysWithDis}.
\begin{lem}\label{lem:RkNonlinear}
    Assume $\Rc_1^\Dc(\Fc)\subseteq\Fc\subseteq\Xc$ and that $\fb(\xb,\ub)-\fb_{\text{PWA}}(\xb,\ub)\in\Dc^{(i)}$ holds for all $\xb\in\Xc$ and for all $\ub\in\Uc$. Then $\Rc_k(\Fc)\subseteq\Rc_k^\Dc(\Fc)$ for all $k\in\N_0$.
\end{lem}

Combining this lemma with the results stated in the sections~\ref{sec:SafeSet} and \ref{sec:TerminalSet} shows that the nonlinear system \eqref{eq:f} is uniformly ultimately bounded (UUB) in the sense of \cite[Def.~2.4]{Blanchini1994}: 
A system is denoted as ultimately bounded in the C-set $\Tc$ (i.e., a convex and compact set containing the origin in its interior), uniformly in $\Fc$, if for every initial condition $\xb(0)\in\Fc$, there exits a $k^\ast(\xb(0))$ such that $\xb(k)\in\Tc$ holds for all $k\geq k^\ast(\xb(0))$.
\begin{thm}\label{thm:UUB}
    Let $\Fc_{\text{max}}$ and $\overline{\Rc}_{\text{min}}$ by defined as in Proposition \ref{prop:Fmax} and \eqref{eq:RminOver}, respectively and assume that $\fb(\xb,\ub)-\fb_{\text{PWA}}(\xb,\ub)\in\Dc^{(i)}$ holds for all $\xb\in\Xc$ and for all $\ub\in\Uc$. Then, the system \eqref{eq:f} with $\ub(k)=\Phib(\xb(k))$ is ultimately bounded in $\overline{\Rc}_{\text{min}}$, uniformly in $\Fc_{\text{max}}$. Moreover, $\xb(k)\in\Xc$ and $\ub(k)=\Phib(\xb(k))\in\Uc$ holds for all $k\in \N_0$.
\end{thm}
\begin{proof}
    Since we have $\Rc_1^\Dc(\Fc_\text{max})\subseteq\Fc_\text{max}\subseteq\Xc$ and $\fb(\xb,\ub)-\fb_{\text{PWA}}(\xb,\ub)\in\Dc^{(i)}$ for all $\xb\in\Xc$ and for all $\ub\in\Uc$ Lemma~\ref{lem:RkNonlinear} applies here. In combination with \eqref{eq:RkSubsetEqRkOver} this results in $\Rc_k(\Fc_\text{max})\subseteq\Rc_k^\Dc(\Fc_\text{max})\subseteq\overline{\Rc}_k^\Dc(\Fc_\text{max})$ for all $k\in\N_0$. Thus we have $\Rc_1(\Fc_\text{max})\subseteq\Rc_1^\Dc(\Fc_\text{max})\subseteq\overline{\Rc}_1^\Dc(\Fc_\text{max})\subseteq\Fc_\text{max}$, i.e. $\Fc_\text{max}$ is PI for the nonlinear system \eqref{eq:f} with NN-based controller \eqref{eq:UOfK}. Therefore, we have $\xb(k)\in\Fc_\text{max}\subseteq\Xc$ for all $k\in\N_0$ and by assumption $\Phib(\xb(k))\in\Uc$ for all $k\in\N_0$. The nonlinear system with \eqref{eq:UOfK} will further converge to $\overline{\Rc}_{\text{min}}$ for all $\xb(0)\in\Fc_{\text{max}}$, since we have $\xb(k)\in\Rc_{\overline{k}}(\Fc_{\text{max}})\subseteq\Rc^\Dc_{\overline{k}}(\Fc_{\text{max}})\subseteq\overline{\Rc}^\Dc_{\overline{k}}(\Fc_{\text{max}})=\overline{\Rc}_{\text{min}}$ for all $k\geq\overline{k}$ and $\xb(0)\in\Fc_{\text{max}}$. This proves that the system \eqref{eq:f} with $\ub(k)=\Phib(\xb(k))$ is ultimately bounded in $\overline{\Rc}_{\text{min}}$, uniformly in $\Fc_{\text{max}}$.
\end{proof}

\section{Case studies}\label{sec:CaseStudy}

In the case study, we consider two different cases. The first case is a PWA system with bounded additive disturbance controlled by an NN-based controller that approximates the MPC control law of the undisturbed system. In the second case study, we consider a nonlinear system approximated by a PWA system.

\subsection{PWA system with additive disturbance}

\begin{figure}[h!]
    \centering
    \includegraphics[trim={0cm 0cm 0cm 0cm},clip,scale=0.5]{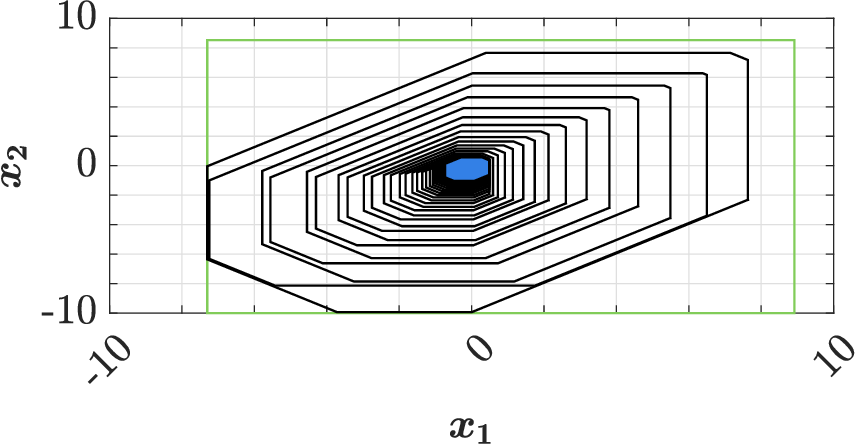}
    \caption{Illustration of the computation of the set $\overline{\Rc}_{\text{min}}=\overline{\Rc}^\Dc_{\overline{k}}(\Fc_{\text{max}})$ in blue starting from the set $\Fc_{\text{max}}$ in green. The black sets illustrate the shrinking sequence of sets $\overline{\Rc}^\Dc_{i}(\Fc_{\text{max}})$ with $i\in\{1,\dots\overline{k}-1\}$ during the computation of $\overline{\Rc}_{\text{min}}$. The computation terminates with the set $\overline{\Rc}^\Dc_{\overline{k}}(\Fc_{\text{max}})=\overline{\Rc}_{\text{min}}$ when \eqref{eq:TerminationFmin} holds.}
    \label{fig:Ri}
\end{figure}

Consider the OCP \cite[Eq.~2]{Teichrib2024} for a nominal PWA system, i.e, with $\db(k)=\zerob$ for all $k\in\N_0$ of the form \eqref{eq:SysWithDis} with 
\begin{align*}
    \Ab^{(1)}&=\begin{pmatrix}
        -0.04 & -0.461 \\
        -0.139 & 0.341
    \end{pmatrix}, 
    \Ab^{(2)}=\begin{pmatrix}
        0.936 & 0.323 \\
        0.788 & -0.049
    \end{pmatrix}, \\
    \Ab^{(3)}&=\begin{pmatrix}
        -0.857 & 0.815 \\
        0.491 & 0.62
    \end{pmatrix},
        \Ab^{(4)}=\begin{pmatrix}
        -0.022 & 0.644 \\
        0.758 & 0.271
    \end{pmatrix},\\
    \Bb^{(1)}&=\Bb^{(2)}=\Bb^{(3)}=\Bb^{(4)}=\begin{pmatrix}
        1 & 0
    \end{pmatrix}^\top,\\
    \Hb^{(1)}&=\begin{pmatrix}
        -1 & 0 \\ 0 & -1
    \end{pmatrix},
    \Hb^{(2)}=\begin{pmatrix}
        -1 & 0 \\ 0 & 1
    \end{pmatrix},
\end{align*}
$\Hb^{(3)}=-\Hb^{(1)},\Hb^{(4)}=-\Hb^{(2)}$, $\hb^{(1)}=\hb^{(2)}=\hb^{(3)}=\hb^{(4)}=\zerob$, $\pb^{(1)}=\pb^{(2)}=\pb^{(3)}=\pb^{(4)}=\zerob$, $\Qb=\Pb=\Ib$, $\Rb=1$, and $N=10$ from \cite[Sec.~7]{Mignone2000} with the additional constraints $\norm{\xb(k)}_\infty\leq 10$ for all $k\in\{0,\dots,N\}$ and $|u(k)|\leq 1$ for all $k\in\{0,\dots,N-1\}$. We solved the OCP for $1000$ randomly sampled $\xb\in\Xc$ to generate a data set with the feasible points $(\xb^{(i)},u^\ast(\xb^{(i)}))$, where $u^\ast(\xb^{(i)})$ is the first element of the optimal input sequence. This data set is used to train a $3\times 3$ maxout NN, i.e., $\ell=3$ and $p_i=w_i=3$ for all $i\in\{1,2,3\}$. Afterwards, we analyzed the PWA system with NN-based controller and an additive disturbance of $\norm{\db(k)}_\infty\leq 0.15$, i.e, $\Dc^{(i)}=\{\db\in\R^n \ | \ \norm{\db}_\infty\leq 0.15\}$ for $i\in\{1,\dots,4\}$ by computing the sets $\Fc_{\text{max}}$ from Proposition~\ref{prop:Fmax} and $\overline{\Rc}_{\text{min}}$ \eqref{eq:RminOver} with $\overline{\epsilon}=10^{-3}$. The computation of $\Fc_{\text{max}}$ terminates after the first iteration, i.e., $\Fc_{\text{max}}=\Fc_1$, where we choose $\Cb=(\Ib \ -\Ib)$ for \eqref{eq:Rover1}. The computation of $\overline{\Rc}_{\text{min}}$ with $\Cb_i=\begin{pmatrix} \cos(\nicefrac{\pi}{4}(i-1)) & \sin(\nicefrac{\pi}{4}(i-1)) \end{pmatrix}$ for all $i\in\{1,\dots,8\}$ can be seen in Figure~\ref{fig:Ri}. The black sets illustrate how the sequence of sets $\overline{\Rc}^\Dc_i(\Fc_{\text{max}})$ with $i\in\{1,\dots,\overline{k}\}$ shrinks until \eqref{eq:TerminationFmin} holds and the computation terminates with $\overline{\Rc}^\Dc_{\overline{k}}(\Fc_{\text{max}})=\overline{\Rc}_{\text{min}}$ (blue set) with $\overline{k}=51$. 

\begin{figure}[h!]
    \centering
    \includegraphics[trim={0cm 0cm 0cm 0cm},clip,scale=0.5]{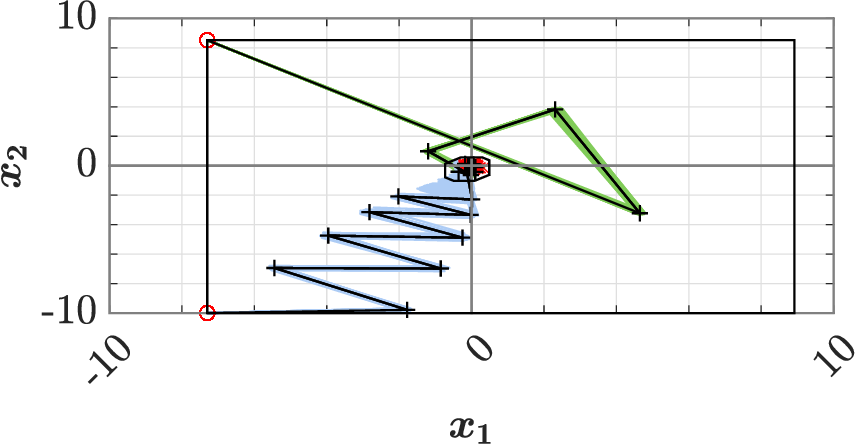}
    \caption{The green and blue “tubes” illustrate in each case $100$ trajectories of the closed-loop system with a random additive disturbance $\norm{\db(k)}_\infty\leq0.15$ in every time step. The black lines represent trajectories of the nominal system (i.e. $\db(k)=\zerob$). The small set around the origin is $\overline{\Rc}_{\text{min}}$ and the box $[-7.3,8.91]\times[-10,8.52]$ is the set $\Fc_{\text{max}}$. The thick grey lines indicate the regions of the PWA system.}
    \label{fig:PWA}
\end{figure}

Since we have \eqref{eq:StabCondFmin} for $\alpha=1+9\times10^{-7}$, Theorem~\ref{thm:BoundedStab} applies here. This means that all trajectories of the PWA system with disturbance starting in $\Fc_{\text{max}}$ remain in that set and enter the stable set $\Fc_{\text{min}}$ after a finite number of time steps. This is illustrated in Figure~\ref{fig:PWA}, where the green and blue “tubes” represent $100$ trajectories with random disturbance for each of the two initial states $\xb(0)=(-7.30 \ 8.52)^\top$ and $\xb(0)=(-7.30 \ -10)^\top$, respectively. As apparent from the figure, even with disturbance, the state constraints are satisfied, and all simulated trajectories enter $\overline{\Rc}_{\text{min}}$ (small black set). Moreover, we can observe an interesting phenomenon. For some blue trajectories, the additive disturbance causes them to be in a different region of the PWA function than the nominal system at the respective time step. This leads to the significant deviation between the nominal trajectory (black) and disturbed blue trajectories shortly before the set $\overline{\Rc}_{\text{min}}$. However, since the computation of the RPI sets $\Fc_{\text{max}}$ and $\overline{\Rc}_{\text{min}}$ is based on reachable sets (cf. Figure~\ref{fig:Ri}) of the disturbed PWA system, these cases are included in the analysis.

\subsection{PWA approximation of a nonlinear system}

\begin{figure}[h!]
    \centering
    \includegraphics[trim={0cm 0cm 0cm 0cm},clip,scale=0.5]{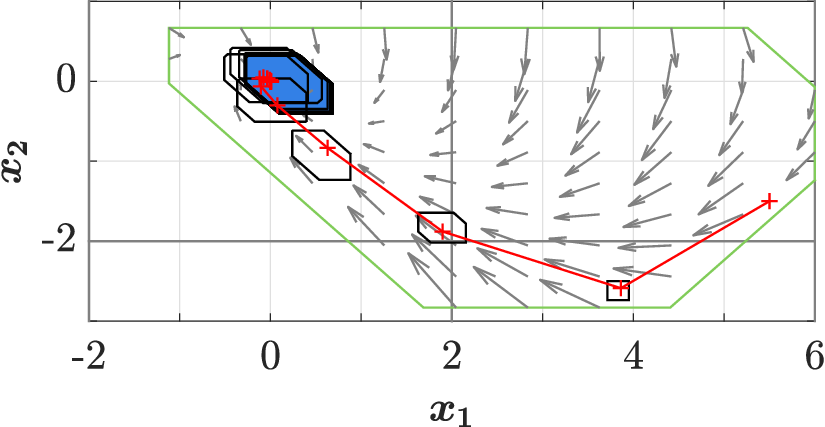}
    \caption{Nonlinear double integrator with an NN-based controller. The grey arrows represent the evolution of the closed-loop system within the set $\Fc_{\text{max}}$ (green set). The red line represents a trajectory $\xb(0),\dots,\xb(\overline{k})$ of the system \eqref{eq:NL_DI} with \eqref{eq:UOfK} starting at $\xb(0)=(5.5 \ -1.5)^\top$. The small sets around the states $\xb(i)$ are the reachable sets $\overline{\Rc}_i^\Dc\big( \{ \xb \in \R^n \ | \ \xb = (5.5 \ -1.5)^\top \} \big)$ with $i\in\{1,\dots,\overline{k}\}$ in which the disturbed PWA system (which is an approximation of \eqref{eq:NL_DI}) and thus the nonlinear system is guaranteed to be in time step $i$. Thick grey lines represent the regions of the PWA system.} 
    \label{fig:NL_DI}
\end{figure}

We consider a system of the form \eqref{eq:f} with
\begin{equation}\label{eq:NL_DI}
    \fb(\xb,u)\!=\!\begin{pmatrix}
        1 & 1 \\
        0 & 1
    \end{pmatrix} \xb\!+\! 
    \begin{pmatrix}
        0.5 \\
        1
    \end{pmatrix} u\!+\! 
    \begin{pmatrix}
        0.025 \\
        0.025
    \end{pmatrix} \xb^\top \xb,
\end{equation}
which is the nonlinear double integrator from \cite[Sec.~6]{Lazar2008}. We solved the OCP from \cite[Sec.~IV]{Lueken2024} with the constraints $\norm{\xb(k)}_\infty\leq6$ and $|u(k)|\leq2$ for $900$ initial values $\xb=\xb(0)\in\Xc$ on a regular grid of size $30\times30$ with $\xb_1$ and $\xb_2$ from $-6$ to $6$ to generate a data set with feasible points $(\xb^{(i)},u^\ast(\xb^{(i)}))$ where $u^\ast(\xb^{(i)})$ is the first element of the optimal input sequence. The data set is used to train a $5\times5$ maxout NN, i.e., $\ell=5$ $p_i=w_i=5$ for all $i\in\{1,\dots,5\}$. For analyzing the closed-loop system with the methods from \ref{sec:AnalysisOfNLSys}, we approximate the system dynamics of the nonlinear double integrator by a PWA function $\fb_{\text{PWA}}(\xb,u):\R^{n+m}\rightarrow\R^n$ with $9$ regions. The $9$ regions divide the state space into a regular chessboard pattern with a side length of $4$, i.e., $\Pc^{(i)}=\{ (\xb^\top \ u)^\top \ | \ \norm{\xb}_\infty\leq2 \}\oplus\{ (\xb^\top \ u)^\top \ | \ \xb=(k \ l)^\top \}$ where the tuple $(k \ l)$ takes the $9$ values from $\{-4,0,4\}\times\{-4,0,4\}$. In each region, we performed a least squares fit of the nonlinear function \eqref{eq:NL_DI} to compute the parameters $\Ab^{(i)}$, $\Bb^{(i)}$, and $\pb^{(i)}$ of the PWA approximation \eqref{eq:SysWithDis}. Which results in an approximation with a maximum error of $0.1186$, i.e., $\norm{\fb(\xb,u)-\fb_{\text{PWA}}(\xb,u)}_\infty\leq0.1186$ for all $(\xb^\top \ u)^\top\in\Xc\times\Uc$. Thus we choose $\Dc^{(i)}=\{\db \in \R^n \ | \ \norm{\db}_\infty \leq 0.1186 \}$ for all $i\in\{1,\dots,9\}$. Since this leads to a PWA approximation with $\fb(\xb,u)-\fb_{\text{PWA}}(\xb,u)\in\Dc^{(i)}$ for all $(\xb^\top \ u)^\top\in\Xc\times\Uc$, Theorem~\ref{thm:UUB} applies here. For the PWA system we computed the sets $\Fc_{\text{max}}$ and $\overline{\Rc}_{\text{min}}$ with $\Cb_i=\begin{pmatrix} \cos(\nicefrac{\pi}{4}(i-1)) & \sin(\nicefrac{\pi}{4}(i-1)) \end{pmatrix}$ for all $i\in\{1,\dots,8\}$ according to Proposition~\ref{prop:Fmax} and \eqref{eq:RminOver}. Where the computation of $\Fc_{\text{max}}$ terminates after $19$ iterations and the computation of $\overline{\Rc}_{\text{min}}$ with $\overline{\epsilon}=10^{-3}$ after $\overline{k}=20$ iterations. The validity of Theorem~\ref{thm:UUB} can be observed in Figure~\ref{fig:NL_DI}. As apparent from the grey arrows, representing the evolution of the nonlinear system \eqref{eq:NL_DI} with NN-based controller \eqref{eq:UOfK}, and the red example trajectory, all trajectories starting in $\Fc_{\text{max}}\subseteq\Xc$ (green set) remain in that set and enter and remain in $\overline{\Rc}_{\text{min}}$ (blue set) after a finite number of time steps. 

\section{Conclusions}\label{sec:Conclusions}

We presented an extension of MI-based analysis for PWA systems with NN-based controllers to the case where the PWA system is assumed to be affected by a bounded additive disturbance. The inclusion of disturbances allows to certify stability and constraint satisfaction for PWA systems with bounded additive disturbances and NN-based controllers (cf. Theorem~\ref{thm:BoundedStab}) and UUB of nonlinear systems approximated by PWA systems (cf. Theorem~\ref{thm:UUB}).

\section*{Appendix}

\subsubsection*{Proof of Lemma~\ref{lem:MI_PWA_with_dis}:} For $\Xb_{K+1},\Ub_{K},\Db_{K},\tilde{\Xb}_{K}^{(1)},...,\tilde{\Xb}_{K}^{(s)}$ subject to \eqref{eq:gammaForR}, \eqref{eq:MI_PWA_xTilde}, \eqref{eq:MI_PWA_sumXTilde}, and \eqref{eq:MI_PWA_ABpd} we have $\xb(k+1)=f_{\text{PWA}}(\xb(k),\ub(k))+\db(k)$ for all $k\in\{1,\dots,K-1\}$ and for all $\xb(0)\in\R^n$, according to \cite[Sec.~3.1]{Bemporad1999}. Since the relation \eqref{eq:GammaXinP} holds for \eqref{eq:gammaForR} we further infer that $\db(k)\in\Dc^{(i)}$ \eqref{eq:dInDLemma} is equivalent to \eqref{eq:MI_PWA_Hd}. Moreover, with \eqref{eq:MI_Constraints_NN} and \eqref{eq:MIConQEqhatX} we have $\Phib(\qb^{(0)}(k))=\Phib(\xb(k))=\Wb^{(\ell+1)}\qb^{(\ell)}(k)+\bb^{(\ell+1)}$ according to \cite[Lem.~2]{Teichrib2023Ifac}. Combining this with \eqref{eq:hatUEqWqplusB} finally results in $\xb(k+1)=f_{\text{PWA}}(\xb(k),\Phib(\xb(k)))+\db(k)$ \eqref{eq:CLSystem} for all $k\in\{1,\dots,K-1\}$ and for all $\xb(0)\in\R^n$, which completes the proof. \hfill $\blacksquare$

\subsubsection*{Proof of Lemma~\ref{lem:suppF_Rk}:} According to Lemma~\ref{lem:MI_PWA_with_dis} the constraints \eqref{eq:MIConClosedLoop1}--\eqref{eq:hatUEqWqplusB} can be replaced by \eqref{eq:PWAConstraints}. Thus, the MILP \eqref{eq:MILPReachableSets} becomes
\vspace{-1mm}
\begin{subequations}\label{eq:MILPReachSetProof}
\begin{align*}
    c_k^\ast(\vb,\Xc)&=\max_{\Xb_{k+1},\Db_k} \vb\xb(k) \\
    \nonumber
    \text{s.t.} \quad &\xb(0)\in\Xc=\Rc_0(\Xc) \\
    \nonumber
    \xb(j&+1)=\Ab^{(i)}\xb(j)+\Bb^{(i)}\Phib(\xb(j))+\pb^{(i)}+\db(j)\\
    \nonumber
    &\db(j)\in\Dc(\xb(j))
\end{align*}
\end{subequations}
for all $j\in\{0,\dots,k-1\}$. By using the definition \eqref{eq:kStepReachSet}, we can replace the constraints for $j=0$ and $\xb(0)\in\Rc_0(\Xc)$ by $\xb(1)\in\Rc^\Dc_1(\Xc)$. Then we can replace the constraints for $j=1$ and $\xb(1)\in\Rc_1(\Xc)$ by $\xb(2)\in\Rc^\Dc_2(\Xc)$. Continuing in this way, we can successively replace all constraints until only $\xb(k)\in\Rc^\Dc_k(\Xc)$ remains. This finally results in 
\begin{equation*}
    c_k^\ast(\vb,\Xc)=\max_{\xb(k)} \vb\xb(k) \quad \text{s.t.} \quad \xb(k)\in\Rc^\Dc_k(\Xc),
\end{equation*}
which is by definition the support function of $\Rc^\Dc_k(\Xc)$ evaluated for $\vb$, i.e., $c_k^\ast(\vb,\Xc)=\hb_{\Rc^\Dc_k(\Xc)}(\vb)$. \hfill $\blacksquare$

\subsection*{Validity of Lemma~\ref{lem:RkNonlinear}}
We first prove the following intermediate result.
    \begin{lem}\label{lem:image_f<=image_fPWAd}
        Let $\fb(\Xc,\Uc)$ and $\fb^\Dc_{\text{PWA}}(\Xc,\Uc)$ be defined as $\fb(\Xc,\Uc)\!:=\!\{\xb^+\!\in\!\R^n \ | \ \xb^+\!=\!\fb(\xb,\ub),\xb \in \Xc,\ub \in \Uc \}$ and $\fb^\Dc_{\text{PWA}}(\Xc,\Uc):=\{\xb^+\in\R^n \ | \ \xb^+=\fb_{\text{PWA}}(\xb,\ub)+\db,\xb \in \Xc,\ub \in \Uc ,\db\in\Dc^{(i)} \}$, respectively. Further assume that $f(\xb,\ub)-\fb_{\text{PWA}}(\xb,\ub)\in\Dc^{(i)}$ for all $\xb\in\Xc$ and for all $\ub\in\Uc$. Then, $\fb(\Fc,\Gc)\subseteq\fb^\Dc_{\text{PWA}}(\Fc,\Gc)$ for $\Fc\subseteq\Xc$ and $\Gc\subseteq\Uc$.  
    \end{lem}
    \begin{proof}
        We need to prove that if $\xb^+\in\fb(\Fc,\Gc)$ then $\xb^+\in\fb^\Dc_{\text{PWA}}(\Fc,\Gc)$. For $\xb^+\in\fb(\Fc,\Gc)$ we have 
    \begin{equation}\label{eq:CondImageF}
        \xb^+=\fb(\xb,\ub), \ \xb\in\Fc, \ \ub\in\Gc.
    \end{equation}
    By assumption there exist a $\db=f(\xb,\ub)-\fb_{\text{PWA}}(\xb,\ub)\in\Dc^{(i)}$ for all $\xb\in\Fc\subseteq\Xc$ and for all $\ub\in\Gc\subseteq\Uc$. Thus \eqref{eq:CondImageF} can be reformulated as follows
    \begin{equation*}
        \xb^+=\fb_{\text{PWA}}(\xb,\ub)+\db, \ \xb\in\Fc, \ \ub\in\Gc, \ \db\in\Dc^{(i)},
    \end{equation*}
    which is exactly the definition of $\fb^\Dc_{\text{PWA}}(\Fc,\Gc)$, i.e., we have $\xb\in\fb^\Dc_{\text{PWA}}(\Fc,\Gc)$ if $\xb\in\fb(\Fc,\Gc)$ for $\Xc\subseteq\Fc$ and $\Gc\subseteq\Uc$. Thus $\fb(\Fc,\Gc)\subseteq\fb^\Dc_{\text{PWA}}(\Fc,\Gc)$ holds for $\Fc\subseteq\Xc$ and $\Gc\subseteq\Uc$.
    \end{proof}
    Now, we use the intermediate result to prove Lemma~\ref{lem:RkNonlinear}.
    \subsubsection*{Proof of Lemma~\ref{lem:RkNonlinear}:} We prove the claim by induction. For $k=1$ we have
    \begin{align*}
        \Rc_1(\Fc)&\!=\!\{ \xb^+\!\in\!\R^n \!\ | \!\ \xb^+\!=\!\fb(\xb,\ub),\xb\!\in\!\Fc,\ub\!\in\!\Uc, \ub\!=\!\Phib(\xb) \} \\
        &=\fb(\Fc,\Gc)
    \end{align*}
    with $\Gc:=\{\ub\in\R^m\ | \ \ub\in\Uc, \ \ub=\Phib(\xb) \}\subseteq\Uc$ and $\Fc\subseteq\Xc$. Then we have according to Lemma~\ref{lem:image_f<=image_fPWAd} 
    \begin{align*}
        \fb(\Fc,\Gc)&\subseteq\fb_{\text{PWA}}^\Dc(\Fc,\Gc) \\
        &=\{\xb^+\in\R^n \ | \ \xb^+=\fb_{\text{PWA}}(\xb,\ub)+\db,\\
        &\hspace{8mm}\xb \!\in\! \Fc,\ub \!\in\! \Uc, \ub=\Phib(\xb) ,\db\!\in\!\Dc^{(i)} \}=\Rc_1^\Dc(\Fc).
    \end{align*}
    In summary $\Rc_1(\Fc)\subseteq\Rc_1^\Dc(\Fc)$ holds for $\Fc\subseteq\Xc$. For $k=i+1$ we have $\Rc_1(\Rc_i(\Fc))\subseteq\Rc_1(\Rc_i^\Dc(\Fc))\subseteq\Rc_1^\Dc(\Rc_i^\Dc(\Fc))=\Rc_{i+1}^\Dc(\Fc)$ and thus $\Rc_{i+1}(\Fc)\subseteq\Rc_{i+1}^\Dc(\Fc)$ if $\Rc_i^\Dc(\Fc)\subseteq \Xc$, which is satisfied since $\Fc\subseteq\Xc$ is RPI by assumption. This proves that $\Rc_k(\Fc)\subseteq\Rc_k^\Dc(\Fc)$ holds for all $k\in\N_0$ and for RPI sets $\Fc\subseteq\Xc$.  \hfill $\blacksquare$

\bibliographystyle{unsrt}        

\end{document}